\documentstyle[12pt,epsf]{article}

\def\be{\begin{equation}}
\def\ee{\end{equation}}
\def\bea{\begin{eqnarray}}
\def\eea{\end{eqnarray}}

\begin{document}
\vspace*{4cm}

\begin{center} 
{\large\bf THE HIGHEST ENERGY NEUTRINOS}\\[4mm]
Enrique Zas \\
{\it Departamento de F\'\i sica de Part\'\i culas,\\
Universidade de Santiago de Compostela, E-15706 Santiago, Spain}
\end{center} 

\abstract{
Some neutrino predictions at the highest energies for a number of 
production mechanisms are comparatively reviewed in the light of future 
projects for neutrino detection.}

\section{Introduction}

High energy neutrino detection is one of the most exciting challenges in particle 
astrophysics because neutrinos provide an alternative view of the Universe. 
Efforts to build such detectors in the forthcoming years \cite{physrep} have 
granted an afternoon session on exprimental high energy neutrinos in this conference. 
Two of these projects, AMANDA and Baikal, are already in operation \cite{here}. 
The ``km$^3$'' initiative, to instrument 1~km$^3$ of water or ice with photodetectors, 
is the natural extention of the lower scale prototypes 
in view of the expected neutrino fluxes \cite{km3case}.  
There is some more motivation in the ``Pierre Auger Project'' to build an air shower detector of 
6000~km$^2$ in search for the highest energy cosmic rays \cite{here}. The project 
is linked to neutrino astronomy in a double way. 
The production mechanism for the highest energy 
cosmic rays must make neutrinos at least in the  
interactions of the cosmic rays with the cosmic microwave 
background and, also, the array itself can be used to detect neutrinos of the highest 
energies \cite{venice}. 

I have been asked to review the possible sources 
of high energy neutrinos which is of course a pretty difficult task to 
do with justice in the light of all the activity in the field and the short space 
available. There is an excellent review provided by Ref. \cite{physrep} 
where more details and complete references can be found, and a  
discussion of neutrino fluxes close to this one in Ref. \cite{protflux}. Keeping this 
in mind I will restrict to some of the production mechanisms that 
predict the highest energy neutrinos. I will discuss their energy shape and comment on the 
plausibility of the mechanisms proposed,   
which is of course pretty subjective, stressing the developments 
in models with neutrinos produced in the jets 
of Active Galaxies.    

\section{Neutrino production by cosmic rays}
   
In the majority of mechanisms most neutrinos arise from the decay of 
charged pions (or kaons), produced in different type of high energy 
particle interactions. 
The pions can be produced in proton-proton or photon-proton interactions 
or alternatively from direct fragmentation of quarks, in the same way they are
produced routinely in electron positron colliders. 
For relativistic pions in flight it can 
be assumed that, on average, each of the four leptons produced in the reaction and 
subsequent muon decay carries one fourth of the parent energy.  

The existence of high energy cosmic rays leaves little doubt about the
actual production of neutrinos in their interactions with 
well understood targets. Atmospheric neutrinos fall in this category 
and are known to within about $10\%$ certainty at energies below 1 PeV 
\cite{physrep}. These neutrinos constitute the 
background for observation of other neutrinos sources. 
Their flux is zenith angle dependent because of the 
competition between interaction and decay of the parent pions. 
The vertical and horizontal 
atmospheric neutrino fluxes are shown in Fig~1A.
At high enough energies the Lorentz expanded lifetime of the pion 
leads to more pion interactions decreasing the relative number of  
neutrinos to their parent pions. This causes that neutrinos from the decays 
of charmed particles (that have considerably 
shorter lifetimes), the ``prompt'' neutrinos, dominate the atmospheric flux 
above some unknown energy 
somewhere above 100 TeV. A typical prompt neutrino prediction \cite{hsvum} is 
illustrated in Fig.~1A. The uncertainty in the prompt component is due to 
the poorly known charm production cross section. 

Cosmic 
rays must also interact with nucleons in the galaxy, such as 
dust, molecular clouds or compact objects like the sun. The interactions with the galactic 
disk matter are most relevant and do not have large uncertainties \cite{protflux}. 
The results \cite{domokos} are also shown in Fig.~1A evidencing that 
these neutrinos dominate the conventional atmospheric flux in the energy 
region where the 
prompt neutrinos are expected. This difficults their possible identification but
it is hoped that the prompt 
neutrinos can be indirectly determined by measuring the atmospheric prompt 
muons which are produced by the same mechanism \cite{hsvum}.  

\begin{figure}[hbt]
\centering
\epsfxsize=6.5in\hspace{0in}
\epsffile{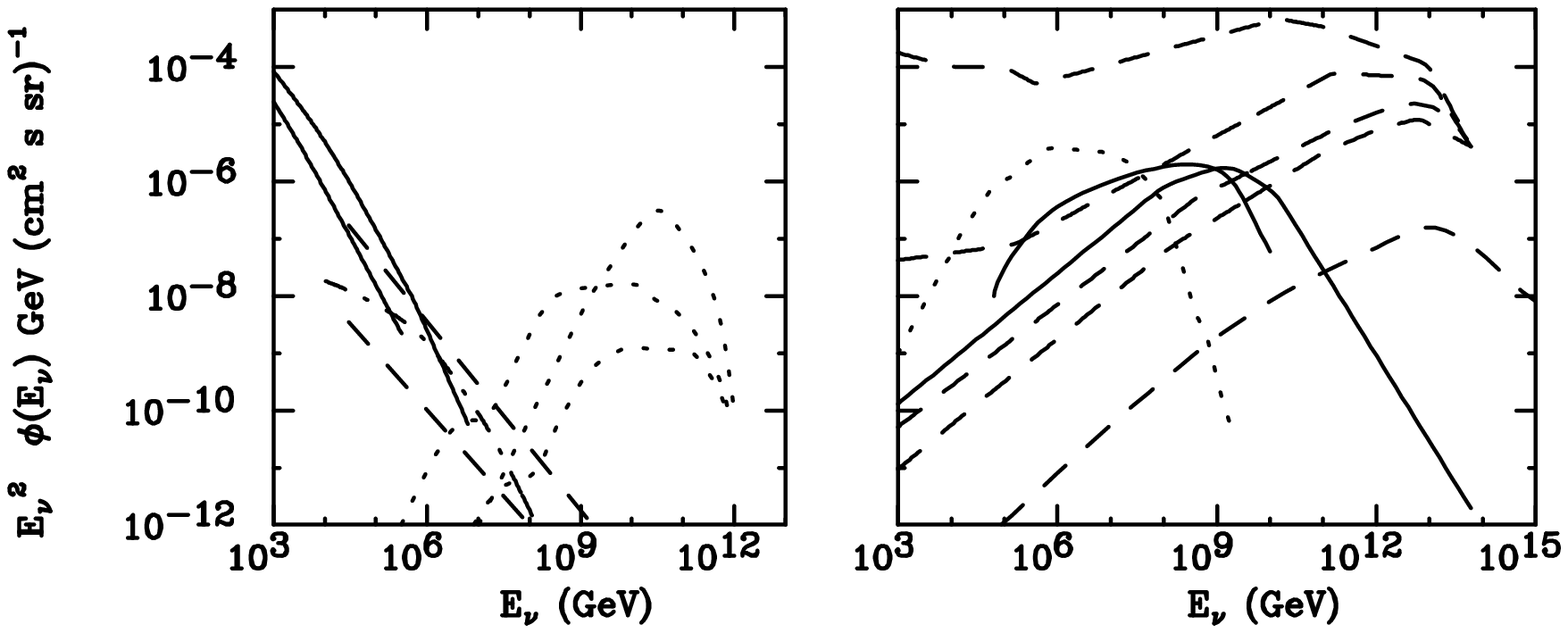} 
\vspace*{-4.in}

\caption{Neutrino flux predictions (from top to bottom where listed): 
5~1A Conventional atmospheric for 90$^0$ and 0$^0$ zenith angles (solid) 
and prompt (dot dashes) [7], from CR interactions with the galactic matter 
for 0$^0$ and 90$^0$ 
galactic latitude 
(dashes) [6] and GZK neutrinos (dots) 
for $z_{ult}=2.2$ [8], 
and $z_{ult}=4$ and $z_{ult}=2$ for 
ref. [9]. 
Fig 1B From AGN jets (solid) (from left to right refs. [16] and 
[15]), 
from AGN cores [12] (dots) and TD decays models with $p=0$, $0.5$, 
$1$ and $1.5$ in ref [18] and for $p=1$ and $m_X=2~10^{25}~$eV
in ref. [19] (dashes).}
\label{fig:fluxes}
\end{figure}

The interactions 
of cosmic rays with the cosmic microwave background is also an 
unavoidable source of neutrinos assuming the higher energy cosmic rays 
are of extragalactic origin and hence universal. This is supported by the non 
observation of cosmic ray anisotropies at high energies.  
Several groups that have
calculated these fluxes\cite{physrep,stecker91,yoshida} and 
their results are within a couple of orders of magnitudes, mostly depending on 
the different assumptions made. I will refer to these as GZK neutrinos to 
stress their relation to the cosmic ray energy cutoff. 
For this calculation the cosmic ray spectrum has to be estimated 
at the production site. This implies extrapolating to 
energies above the maximum currently observed ($\sim3~10^{20}$~eV) and 
making some assumptions about the evolution of cosmic ray luminosity with 
cosmological time.     
The production mechanism has to be integrated over time 
(or redshift) up some earlier epoch ($z_{ult}$) which is expected 
to be provided by the Galaxy formation era ($z_{ult}$=2-4). 
These flux predictions are all fairly flat because the proton photon 
interaction cross section has a threshold behavior at the resonant 
$\Delta$ production. Most neutrinos are produced with energies about 
a factor 20 (see next section) below the Greisen-Zatsepin-Kuz'min 
cutoff energy $\sim 10^{20}~$eV. 
Depending on ($z_{ult}$) the interactions of the highest energy neutrinos 
with the cosmic neutrino background can play a more important role altering 
their shapes in the highest energy region.
Fig.~1A includes some of these calculated fluxes indicating the levels of 
uncertainty. 

Regardless of the uncertainties in the GZK neutrinos, all these mechanisms 
are certain, at least in the sense that if by some means they were found not 
to be there, the hypothetical implications of such non-discovery would have 
a larger impact in physics and/or cosmology than their actual observation. 
 
\section{Neutrino production in objects known to exist}   

I will now consider another category of neutrino fluxes which is plausible 
in the sense that they can be produced in objects that we know exist. Some of 
them can be galactic such as accretion in binary systems, 
supernova remnants, 
but those reaching to highest energies are likely to be extragalactic. 
The most representative are Active Galactic Nuclei (AGN), 
and possibly Gamma Ray Bursts (GRB), although the origin of GRB's is still in debate. 
AGN have also been dedicated a good part of a morning session in this conference. 
The recent detection of GeV gamma rays from over sixty 
AGN by the Compton Gamma Ray Observatory (GRO)\cite{GRO} together with de detection 
of TeV photons from three 
other nearby AGN with the imaging technique in Cherenkov telescopes \cite{weekes}, 
place them at the forefront of particle astrophysics. 
These objects have also been proposed as sites for acceleration of the highest 
energy cosmic rays, as they have physical parameters which are 
dimensionally compatible with high energy cosmic rays. 
For all these reasons I will discuss them in some detail. 

AGN are the most luminous 
objects that we observe. They display remarkable jets that   
stream highly collimated out of their cores to distances of several parsecs. 
They also show inner structure with superluminal motion, which is naturally 
explained by particle flows with bulk relativistic speeds.  
These jets observed in the radio band are very likely 
due to synchrotron emission from electrons that  
are accelerated along the jet axis. 
If protons are accelerated along with the electrons as some authors 
claim, then neutrino production is unavoidable because 
of photoproduction of pions in the ambient radiation field which is 
close to the Eddington limit. Earlier models of such fluxes 
considered their production in AGN cores \cite{stecker95} but the 
recent identification of all gamma ray detections with blazars \cite{mattox},
believed to be AGN with their jets pointing towards us \cite{padovani}, has 
shifted the interest to models in which protons are accelerated in the jets 
\cite{mannheim95,protheroe96}. In these models the neutrinos are Lorentz boosted 
to energies higher than in AGN cores, what has important 
implications for their detection. 

The models can be dimensionally explained with 3 simple assumptions \cite{apjzas}: 
protons are 
accelerated in the jets with an $E^{-2}$ spectrum as expected in shock 
acceleration, the maximum energy for the protons is $10^{20}$~eV and, finally, the 
target photon density behaves as a negative power law $E^{-\alpha}$ (for  
AGN in the broad infrared to X-ray band $\alpha$ is typically around 1).  

It can be shown by 
simple energetics of the photopion production that the ratio of 
neutrino to photon luminosities is roughly $3/13$. The result is obtained using 
a cross section for $\pi^0$ production twice that for 
$\pi^+$, assuming each neutrino has exactly one fourth of the $\pi^+$ energy
 and adding a small correction for pair production
\cite{physrep}. The neutrino energy flux can be obtained scaling the measured GeV to 
TeV gamma ray energy flux for Markarian 421, $J_{\gamma} \sim 5~10^{-10} $~TeV~cm$^{-2}$~s$^{-1}$ 
with this ratio to get $J_{\nu} \sim 10^{-10} $~TeV~cm$^{-2}$~s$^{-1}$. 
Mrk 421 is a nearby blazar which has been well established by GRO and by two Cherenkov 
telescopes. The 
shape of the neutrino spectrum here is also finally obtained by the threshold behavior of
the photoproduction cross section at the $\Delta$ resonance. 
For a given proton energy $E_p$ the required energy of the target photon
for resonance scales as $E_p^{-1}$. Combining the proton 
spectrum and the target photon density spectrum at resonance gives 
a power law $E_{\nu}^{-2+\alpha}$ with maximum neutrino 
energy of $2~10^{18}~eV \sim 0.25 <x_F>~E_p^{max}$ (where $<x_F>$=0.2 is the average 
Feynman-$x$ for photopion production at resonance). 
Provided $\alpha > 0$, the total energy flux, the spectral index and the maximum energy 
determine the flux uniquely 
because then the energy integral is insensitive to the lower limit. 
The shape of the spectrum obtained is very close to that predicting in the two 
alternative models for acceleration in AGN jets. It is a straightforward matter to 
rescale the flux with some factor of order 100~sr$^{-1}$ corresponding to the 
equivalent number of Mrk421 flux-like AGN per stereoradian\cite{apjzas}, to get also the 
order of the normalization for the diffuse neutrino flux from all AGN. 
The exercise stresses the important assumptions in
the models and explains the overall shift of energy to the $10^{18}$~eV region as
illustrated in Fig.~1B where the two predictions for acceleration of protons in 
jets are compared to neutrinos from their cores.  
 
\section{Exotic neutrino sources} 

Lastly there is a third category of more exotic sources whose existence has 
only been postulated on theoretical grounds. Such is the case of Primordial 
Black Holes, decays of topological defects or WIMP 
annihilation. Topological defect (TD) scenarios arise in grand unified theories
of particle interactions with spontaneous symmetry breakdown. They are naturally 
formed as some field vacuum goes through a phase transition to a new degenerate 
vacuum as the Universe cools down in its expansion. Different regions of space
go to different vacuua and the net distribution may 
evolve later into a vacuum field with non-trivial 
topology, surrounding a point (monopole), a line (string) or a surface 
(domain wall). These cosmological 
objects accumulate energy and when they interact with themselves or with 
other objects of their own nature, they annihilate liberating large amounts of 
energy in the form of $X$ particles, the Gauge bosons of the underlying grand 
unified theory. TD scenarios have been recently heavily
discussed as the possible origin of the highest energy cosmic rays.  
Several authors have normalized the defect abundances to cosmic or gamma 
ray measurements and bounds \cite{bahtta92,sigl,protstan}. Such a mechanism avoids 
the conceptual difficulties involved in accelerating protons or nuclei 
in our venicity to energies above the Greisen Zatsepin Kuz'min cutoff. 

The models are however very uncertain because they are significantly affected
by a variety of parameters besides the normalization itself. In general 
they extend to very high energies dictated by the mass of the $X$ particles 
expected to be of order $10^{14}-10^{16}~$GeV. The shape of the fluxes 
predicted are very flat and are somewhat different depending on the 
behavior of the time evolution of the effective 
injection rate of $X$ particles. This is usually parameterized as $t^{p-4}$ 
with $p=0$ for superconducting cosmic strings, $p=1$ for monopoles and cosmic 
strings and $p=2$ for constant injection rates in comoving volume \cite{bahtta92}. 
The main uncertainty in the neutrino spectrum shape is due to 
the fragmentation function assumed which is 
used with large extrapolations. Moreover the normalization to cosmic or 
gamma rays is also subject to uncertainties due to the interactions of the cosmic 
rays in their propagation, mostly with the poorly known extragalactig $B$ fields 
\cite{protstan}. Fig.~1B illustrates some of the produced neutrino 
fluxes by different authors and for different assumptions.

\section{Experiment: present and future}

There are already some experimental results in the form of upper bounds provided
by three types of experiments. One is from underground muon detectors, of which 
Fr\`ejus provides the most stringent limit \cite{frejus}, 
the other two are from Extensive Air Shower detectors, particle detector arrays such as 
AKENO and EAS-TOP and a fluorescence light detector, Fly's Eye. Their results are 
not straightforward to convert to bounds on differential neutrino spectra 
because the conversion involves an assumption on the shape of the neutrino spectrum. 
Moreover there are important uncertainties in the high energy neutrino cross sections, 
besides 
the usual experimental uncertainties associated with each of the experiments. Fig.~2
compares atmospheric fluxes, some TD fluxes and fluxes from AGN jets to these bounds. 
Some of the results are clearly in conflict with experiment. Some Superconducting 
Cosmic String models are ruled out by underground detectors and by the muon poor 
horizontal shower bound from AKENO \cite{blanco}. 

\begin{figure}[t]
\centering
\epsfxsize=6in\hspace{0in}\epsffile{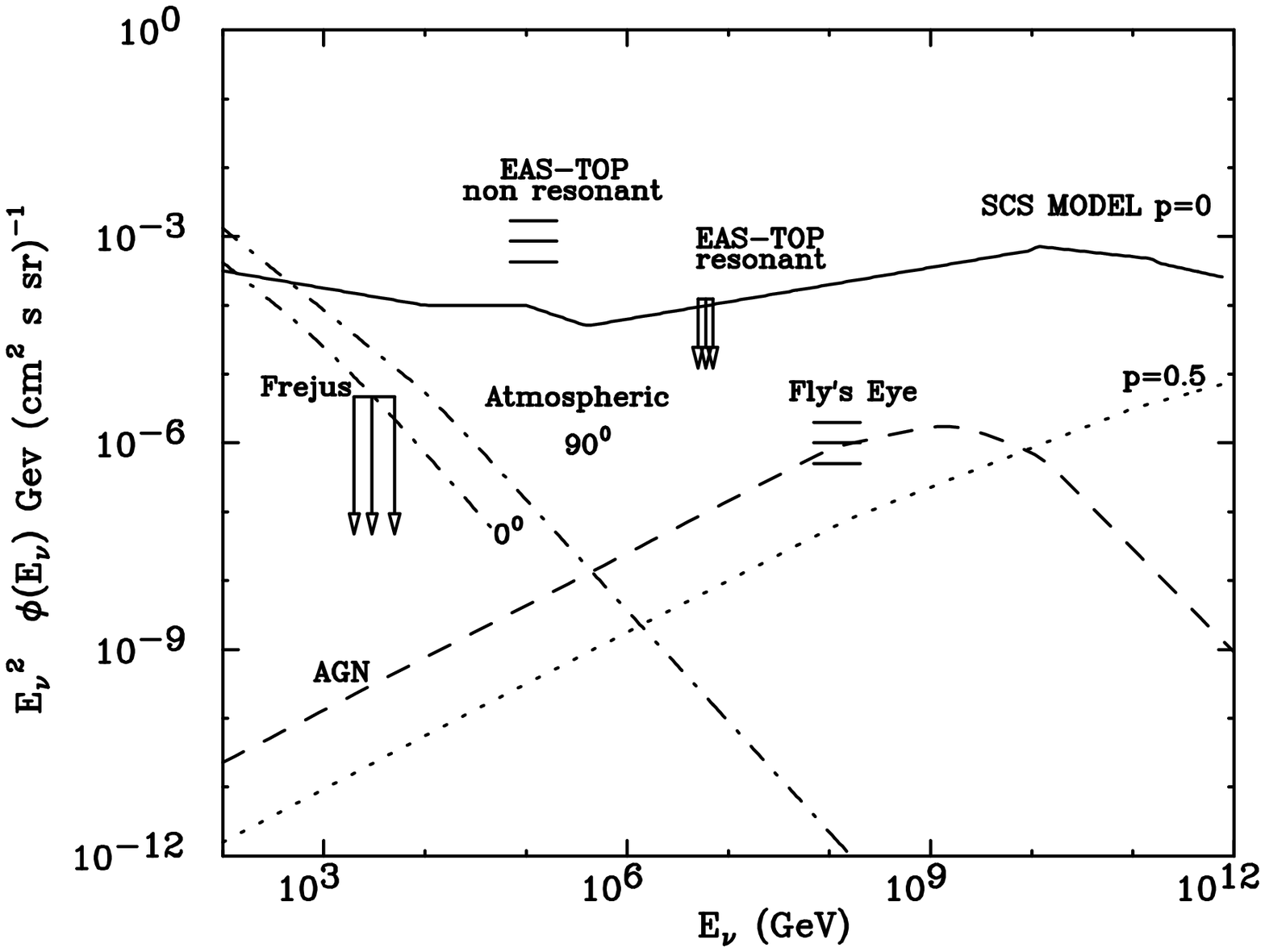} 
\vspace*{-2.in}

\caption{Neutrino flux predictions compared to existing bounds from experiment. 
Parallel lines indicate the uncertainty associated to the spectral index assumed 
for the conversion to a flux bound.}
\label{fig:bounds}
\end{figure}

Our discussion of fluxes has stressed the importance that the highest energy 
neutrinos have for the future of this field, particularly in the light of recent
theoretical developments. Possibly the largest challenge is provided by the
low level of the GZK neutrinos which should however be there. The shift of interest 
to the higher energies has some important implications for detection because of the
rise in the neutrino cross section. At these energies the Earth will be completely 
opaque to neutrinos so underground 
muon detectors will have to look for horizontal showers or vertical downgoing 
showers. Moreover the showers will be in dense media where interesting new effects 
such as the Landau-Pomeran\v cuck-
Migdal will markedly show up and difficult the energy measurements. 
This will certainly affect the optimal separation of their optical modules. A shift
to neutrino energies in the $10^{18}~$eV and above adds considerable more interest 
to the alternative techniques such as the detection of horizontal air showers with 
giant array detectors like the Auger project or the yet unproven detection of the 
coherent radio pulses from the excess charge in the 
showers\cite{zas}. 
In table~1 some preliminary results 
of expected neutrino event rates in the Pierre Auger project for a number of the 
discussed fluxes are reported \cite{venice}. 

\begin{table}[t]
\caption{Expected neutrino event rates in the Pierre Auger Project for several fluxes.}
\vspace{0.4cm}
\begin{center}
\begin{tabular}{|c|c|}
\hline
 {$\nu$~source}  & Range of yearly event rates \\ 
\hline
{AGN-cores} \cite{stecker95} &  0.2-1.5 \\ 
{AGN-jets} \cite{mannheim95} & 2-7 \\
{GZK $z_{ult}=4$} \cite{yoshida} & 0.1-0.4 \\
{TD $p=1.5$} \cite{bahtta92}   &  2-10 \\
\hline
\end{tabular}
\end{center}
\end{table}

Hopefully in the near future we will have some neutrino events. Underground muon detectors
with a 1~km$^2$ surface area will have 
very enhanced acceptances for muon neutrinos because of the long range of the muon 
produced in charged current interactions, and 
can detect neutrinos for energies starting from roughly 100 GeV or so. 
The Pierre Auger Project could have 
an acceptance comparable to 1~km$^3$ for contained events and electron neutrinos. 
Lastly the radio technique if proven to be viable may open new possibilities of 
exploring even larger energies and lower fluxes. 
The complementarity between each of the detector types would
no doubt constrain any hypothetically detected flux and 
allow the extraction of much more precise information. 

\section*{Acknowledgments}

I thank Gonzalo Parente for reading the manuscript and J. Alvarez Mu\~niz, 
J.J. Blanco Pillado, F. Halzen, A. Letessier-Selvon, K.~Mannheim,  
R.~Protheroe, G.~Sigl and R. V\'azquez for discussions. This work was supported 
in part by CICYT (AEN96-1773) and by Xunta de Galicia (XUGA-20604A96).



\end{document}